\documentclass[aps, prl, twocolumn, superscriptaddress, amsmath 
]{revtex4-2}

\usepackage{graphicx}
\usepackage{amsfonts}
\usepackage{amssymb}
\usepackage{amsmath}
\usepackage{lineno}
\usepackage{booktabs}
\usepackage{multirow}

\usepackage[normalem]{ulem}
\usepackage{hyperref}
\hypersetup{
    colorlinks=true,       
    linkcolor=blue,          
    citecolor=blue,        
    filecolor=blue,      
    urlcolor=blue           
}
\usepackage{makecell}
\usepackage{tabularx}
\usepackage{array}

\newcommand{\ket}[1]{\vert#1\rangle}

\newcommand{\be}{\begin{equation}}
\newcommand{\ee}{\end{equation}}
\newcommand{\ba}{\begin{array}}
\newcommand{\ea}{\end{array}}
\newcommand{\bea}{\begin{eqnarray}}
\newcommand{\eea}{\end{eqnarray}}

\def\6{{\langle}}
\def\9{{\rangle}}

\usepackage{xcolor}

\begin{document}

\title{Violation of Bell Inequality with Unentangled Photons}

\author{Kai Wang}
\email{These two authors contributed to this work equally}
\affiliation{National Laboratory of Solid State Microstructures, School of Physics, Collaborative Innovation Center of Advanced Microstructures, Nanjing University, Nanjing 210093, China}
	
\author{Zhaohua Hou}
\email{These two authors contributed to this work equally}
\affiliation{National Laboratory of Solid State Microstructures, School of Physics, Collaborative Innovation Center of Advanced Microstructures, Nanjing University, Nanjing 210093, China}
	
\author{Kaiyi Qian}
\affiliation{National Laboratory of Solid State Microstructures, School of Physics, Collaborative Innovation Center of Advanced Microstructures, Nanjing University, Nanjing 210093, China}

\author{Leizhen Chen}
\affiliation{National Laboratory of Solid State Microstructures, School of Physics, Collaborative Innovation Center of Advanced Microstructures, Nanjing University, Nanjing 210093, China}

\author{Mario Krenn}
\email{mario.krenn@mpl.mpg.de}
\affiliation{Max Planck Institute for the Science of Light (MPL), Erlangen, Germany}
\affiliation{Machine Learning in Science, Department for Computer Science, Faculty of Science, University of Tuebingen.}

\author{Markus Aspelmeyer}
\email{markus.aspelmeyer@univie.ac.at}
\affiliation{Institute for Quantum Optics and Quantum Information, Austrian Academy of Sciences, Boltzmanngasse 3, Vienna A-1090, Austria}
\affiliation{University of Vienna, Faculty of Physics, Vienna Center for Quantum Science and Technology (VCQ), Boltzmanngasse 5, Vienna A-1090, Austria.}

\author{Anton Zeilinger}
\email{anton.zeilinger@univie.ac.at}
\affiliation{Institute for Quantum Optics and Quantum Information, Austrian Academy of Sciences, Boltzmanngasse 3, Vienna A-1090, Austria}
\affiliation{University of Vienna, Faculty of Physics, Vienna Center for Quantum Science and Technology (VCQ), Boltzmanngasse 5, Vienna A-1090, Austria.}

\author{Shining Zhu}
\email{zhusn@nju.edu.cn}
\affiliation{National Laboratory of Solid State Microstructures, School of Physics, Collaborative Innovation Center of Advanced Microstructures, Nanjing University, Nanjing 210093, China}
	
\author{Xiao-Song Ma}
\email{Xiaosong.Ma@nju.edu.cn}
\affiliation{National Laboratory of Solid State Microstructures, School of Physics, Collaborative Innovation Center of Advanced Microstructures, Nanjing University, Nanjing 210093, China}
\affiliation{Hefei National Laboratory, Hefei, 230088, China. }
\affiliation{Synergetic Innovation Center of Quantum Information and Quantum Physics, University of Science and Technology of China, Hefei, Anhui 230026, China.}
	
\date{\today}
	
\begin{abstract}
Violation of local realism via Bell inequality— a profound and counterintuitive manifestation of quantum theory that conflicts with the prediction of local realism — is viewed to be intimately linked with quantum entanglement. Experimental demonstrations of such a phenomenon using quantum entangled states are among the landmark experiments of modern physics and paved the way for quantum technology. Here we report the violation of the Bell inequality that cannot be described by quantum entanglement in the system but arises from quantum indistinguishability by path identity, shown by the multi-photon frustrated interference. By analyzing the measurement of four-photon frustrated interference within the standard Bell-test formalism, we find a violation of Bell inequality by more than four standard deviations. Our work establishes a connection between quantum correlation and quantum indistinguishability, providing insights into the fundamental origin of the counterintuitive characteristics observed in quantum physics.
\end{abstract}

\maketitle
Interference of single quanta epitomizes the counterintuitive character of quantum physics. As Feynman said, the interference of the single quantum “has in it the heart of quantum mechanics” \cite{feynman1965lectures}. In all interference phenomena, there is more than one possibility giving rise to an event (such as detecting a photon at the outputs of an interferometer). Only when these possibilities are indistinguishable, their probability amplitudes are summed up, which leads to quantum interference \cite{mandel1991coherence,RevModPhys.71.s274,RevModPhys.71.s288,zeilinger2005happy}.

By extending interference phenomena from single quanta to multiple quanta, one creates entanglement \cite{PhysRev.47.777,PhysRev.48.696,schrodinger1935gegenwartige}, the coherent superposition of states shared by several quanta. Entanglement allows quantum correlations between many quanta that are stronger than classically possible, which is manifested by multi-quanta interference. Quantum correlations cannot be explained or reproduced by classical physics, as proven by Bell's theorem \cite{PhysicsPhysiqueFizika.1.195,RevModPhys.86.419}. Entanglement is one of the most profound traits of quantum mechanics, and is a key resource for quantum information technology. 

In 1991, Zou, Wang, and Mandel realized a novel type of quantum interference and explored the concept of quantum indistinguishability in a mind-boggling way \cite{PhysRevLett.67.318,greenberger1993multiparticle}. In their work and the later work by Herzog et al. \cite{PhysRevLett.72.629}, two processes of photon-pair generation were arranged such that the paths of the emitted photons are indistinguishable. The path information never exists in the first place due to this indistinguishability of the generation processes. The detection of individual photons carries no information about where they are generated. Consequently, Herzog et al. observed so-called frustrated interference (FI), in which the photon-pair-generation processes can be suppressed or promoted by varying the interferometric phases. This kind of quantum interference due to indistinguishability has been applied to such fields as quantum imaging\cite{lemos2014quantum,yang2023interaction},  spectroscopy \cite{kalashnikov2016infrared}, optical coherence tomography \cite{paterova2018tunable}, quantum state generation and analysis \cite{PhysRevLett.118.080401,su2019versatile,PhysRevLett.130.090202}, microscopy \cite{paterova2020hyperspectral,kviatkovsky2020microscopy}, and quantum holography \cite{topfer2022quantum}. Related quantum interference has also been observed with two weak fields from a local oscillator \cite{PhysRevLett.87.123603}.

However, two-photon FI is an intrinsically local phenomenon \cite{PhysRevLett.73.3041}. This is due to two key factors: First, individual photon counts exhibit interference based on the combined phases of both photons, leading to interference in joint outcomes. This is fundamentally different to entanglement, in which joint outcomes interfere as well, but individual photon counts exhibit no interference. Second, the experimental settings, such as those determining the phases of each photon, are decided inside the backward light cones of the detection events.

Here we show that much richer phenomena appear with correlations between more than two particles, by demonstrating experimental violation of Bell inequality through quantum indistinguishability by path identity, rather than quantum entanglement. The key to these experiments is multi-photon FI, which was recently proposed \cite{gu2019quantum} and demonstrated \cite{feng2023chip,qian2023multiphoton}. In the present work, we demonstrate the violation of a Bell inequality by more than four standard deviations and establish that the observed effect cannot be explained by entanglement, neither at the outcome nor anywhere within the experimental setup. Our result has intriguing consequences for the foundations of quantum information science, and opens up a different perspective on the question regarding the fundamental origin of conflict between quantum mechanics and local realism.

\begin{figure*}
\includegraphics[width=14 cm]{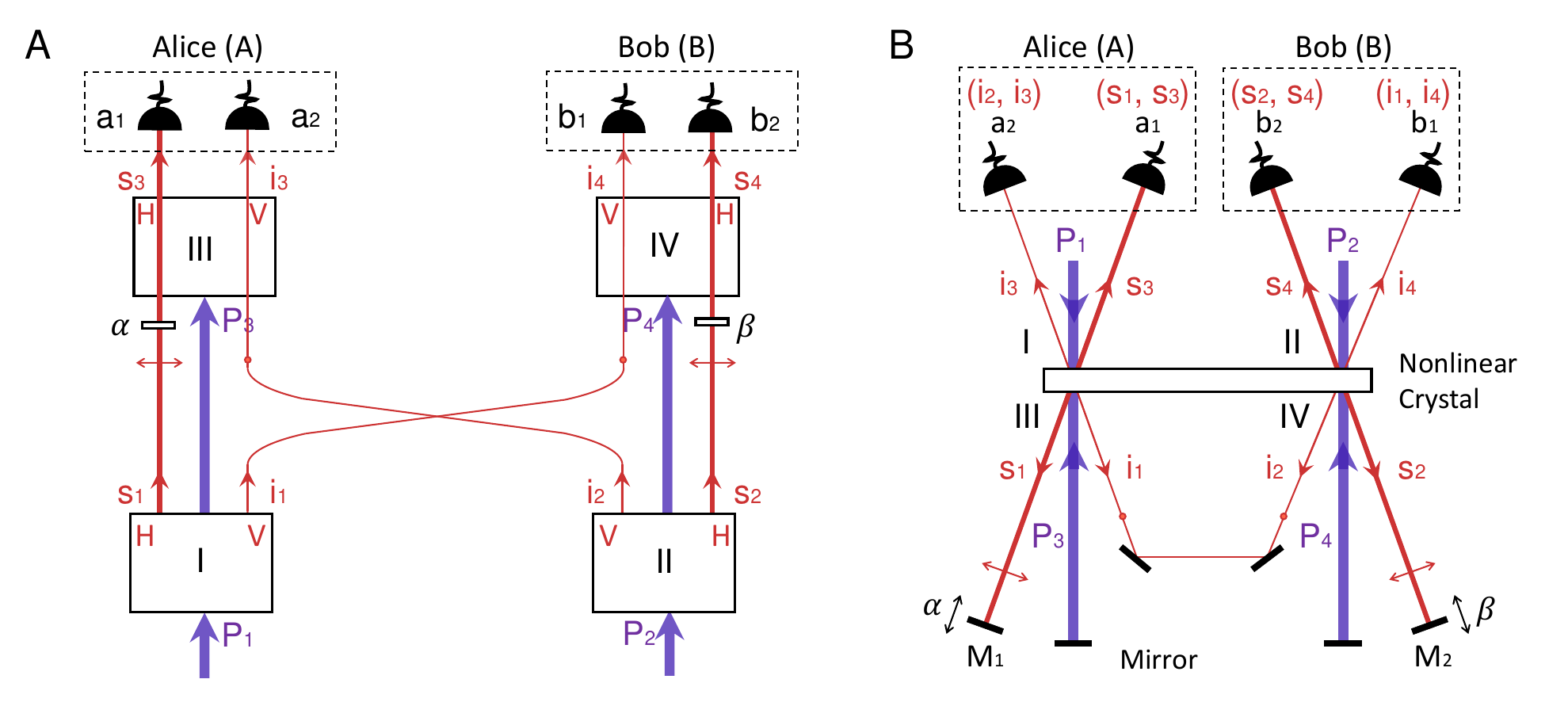}
\caption{Four-photon frustrated interference. (A) Every white square (I, II, III, IV) represents a two-photon source that can generate polarization product states in $\ket{HV}$. The four-photon state we post-selected can either be generated from sources I and II ($s_1$, $i_1$, $i_2$, $s_2$) or that from sources III and IV ($s_3$, $i_4$, $i_3$, $s_4$). When the photons on each individual path are identical in every degree of freedom, they interfere with each other. (B) Simplified scheme of the experimental setup. Pumps P1 and P2 provide the possibility to generate two pairs of photons ($s_1$, $i_1$)I and ($s_2$, $i_2$)II in the form of polarization product states from a nonlinear crystal. The pump beams are then reflected as P3 and P4, providing the possibility to generate another two pairs of photons ($s_3$, $i_3$)III and ($s_4$, $i_4$)IV, respectively. The signal photons (horizontally polarized) $s_1$ and $s_2$ are reflected back to the crystal by two mirrors M1 and M2, respectively, and are aligned with $s_3$ and $s_4$, respectively, to ensure path indistinguishability. Photons $s_1$ and $s_3$ are then in the mode $a_1$, and photons $s_2$ and $s_4$ in the mode $b_2$. The idler photons $i_1$ and $i_2$ (vertically polarized) propagate along the same path but in opposite direction. This is equivalent to swapping their path in Fig. 1A. Alice (Bob) detects photons on the modes $a_1$ and $a_2$ (modes $b_1$ and $b_2$). By moving M1 (M2), we change the measurement settings through controlling the phase $\alpha$ ($\beta$). The phase stability, including the relative phases of four pump beams, signal photons and idler photons, in this four-photon frustrated interferometer is crucial and is passively maintained experimentally. }
\end{figure*}

\section{Experimental Setup}
The scheme of our experiment is shown in Fig. 1A. There are four probabilistic two-photon sources, which are coherently pumped by classical lasers and are labelled as sources I–IV (represented by white squares in Fig. 1A). Each source generates a pair of photons (signal and idler photons) in a polarization product state ($\ket{HV}$, H/V represents horizontal/vertical polarization). The spatial modes of signal photons (H-polarized) from sources I\&III and II\&IV are aligned identically to the paths $a_1$ and $b_2$, shown with thicker red lines in Fig. 1A. The spatial modes of idler photons (V-polarized) from sources I \& II are swapped, and then aligned with those of the idler photons from sources III \& IV, whose paths are identically to $a_2$ and $b_1$, shown with thin red lines in Fig. 1A. The phases of the signal photons at Alice and Bob’s sides, $\alpha$ and $\beta$, are set by the respective phase shifters. The final state of the photons occupying the modes $a_1$, $a_2$, $b_1$, $b_2$ by post-selecting the four-photon term is:
\begin{equation}
  |\psi_{f}\rangle _{a_1a_2b_1b_2}=g^2[e^{i(\alpha +\beta )}|1111\rangle +|1111\rangle ]
\end{equation}
where the numbers represent the photon-number states in the respective modes $a_1$, $a_2$, $b_1$, $b_2$. For the complete form of the output state to the second-order approximation, see Methods. The parameter g is the efficiency of the spontaneous parametric down conversion (SPDC) process, which depends on the pump power and the nonlinearity of the nonlinear crystal. The generated rate of the four-photon-coincidence events by simultaneously detecting $a_1$, $a_2$, $b_1$, $b_2$ (Eq. (1)) is a function of both phases $\alpha$ and $\beta$: 
\begin{equation}
  \mathcal{N} (\alpha ,\beta )=g^4N_0[2+2\cos\mathrm{(}\alpha +\beta )]
\end{equation}
where $N_0$ is the repetition rate of the classical pump light with high intensity.

A simplified schematic of our experimental setup is shown in Fig. 1B. See Fig. A2 in Appendix for the detailed setup. Two parallel and mutually coherent pump beams P1 and P2 (violet arrows) provide the possibility to generate two pairs of photons from crystals I ($s_1$, $i_1$) and II ($s_2$, $i_2$) in the form of polarization product state from the nonlinear crystal. The paths of idlers from sources I ($i_1$) and II ($i_2$) are swapped by routing the same path reversely. The signal photons $s_1$ and $s_2$ are reflected by two mirrors (M1, M2), with which Alice and Bob can change their measurement settings of phases $\alpha$ and $\beta$ by moving their respective mirror. The propagating pumps are reflected back to the crystal (as beams P3 and P4). They provide the possibility to generate another two photon pairs III ($s_3$, $i_3$) and IV ($s_4$, $i_4$). Based on the scheme shown in Fig. 1A, the modes of photons from sources III and IV are aligned with the photons from sources I and II: $s_1$ and $s_3$ occupying mode $a_1$; $i_1$ and $i_4$ occupying mode $b_1$; $i_2$ and $i_3$ occupying mode $a_2$; and $s_2$ and $s_4$ occupying mode $b_2$. We emphasize that photons on the same path are indistinguishable in every degree of freedom. Therefore, when we detect four photons simultaneously on modes $a_1$, $a_2$, $b_1$ and $b_2$, we cannot determine whether they are from sources I and II, or from sources III and IV. Then we obtain the interference of the two possible processes, and the phase-dependent four-fold-coincidence counts, as given by Eq. (2).

Initially, the two photons from the sources in our experiment are entangled in momentum and frequency. We then destroy their momentum entanglement using single-mode fiber coupling, which performs strong projective measurements on the momentum of the photons. By using band-pass filters (~3 nm bandwidth), we also destroy the frequency entanglement between them. Therefore, we cannot utilize these internal entangled degrees of freedom in the Bell-inequality test presented below. 

In our analysis, we represent the simultaneous detection of the two photons ($a_1$, $a_2$ for Alice, and $b_1$, $b_2$ for Bob, see Fig. 1) as outcome +1. To get outcome -1 in the phase setting $\alpha$, Alice sets the phase to an orthogonal base $\alpha+\pi$ and then detects two photons simultaneously. Similarly, Bob sets the phase $\beta+\pi$ to get his -1 outcome. This is based on the cosine-phase dependence with period 2$\pi$ in the frustrated interference of our work, see Fig. A3 in Appendix. We therefore make the following assumptions about the count rate $N(a,b|\alpha,\beta)$ to construct the joint probability: $N(+1, -1|\alpha,\beta)= N(+1, +1|\alpha,\beta+\pi)$; $N(-1, +1|\alpha,\beta)=N(+1, +1|\alpha+\pi,\beta)$; $N(-1, -1|\alpha,\beta)= N(+1, +1|\alpha+\pi,\beta+\pi)$. For details, see Methods. Using this procedure, the probabilities of a (+1, +1) outcome for the setting $(\alpha, \beta)$ is:
\begin{widetext}
\begin{align}
  \nonumber p(+1,+1|\alpha ,\beta )&=\frac{\mathcal{N} (\alpha ,\beta )}{\mathcal{N} (\alpha ,\beta )+\mathcal{N} (\alpha +\pi ,\beta )+\mathcal{N} (\alpha ,\beta +\pi )+\mathcal{N} (\alpha +\pi ,\beta +\pi )}\\
  &=\frac{1}{4}+\frac{1}{4}\cos\mathrm{(}\alpha +\beta ),
\end{align}
\end{widetext}
\begin{figure*}
\includegraphics[width=14 cm]{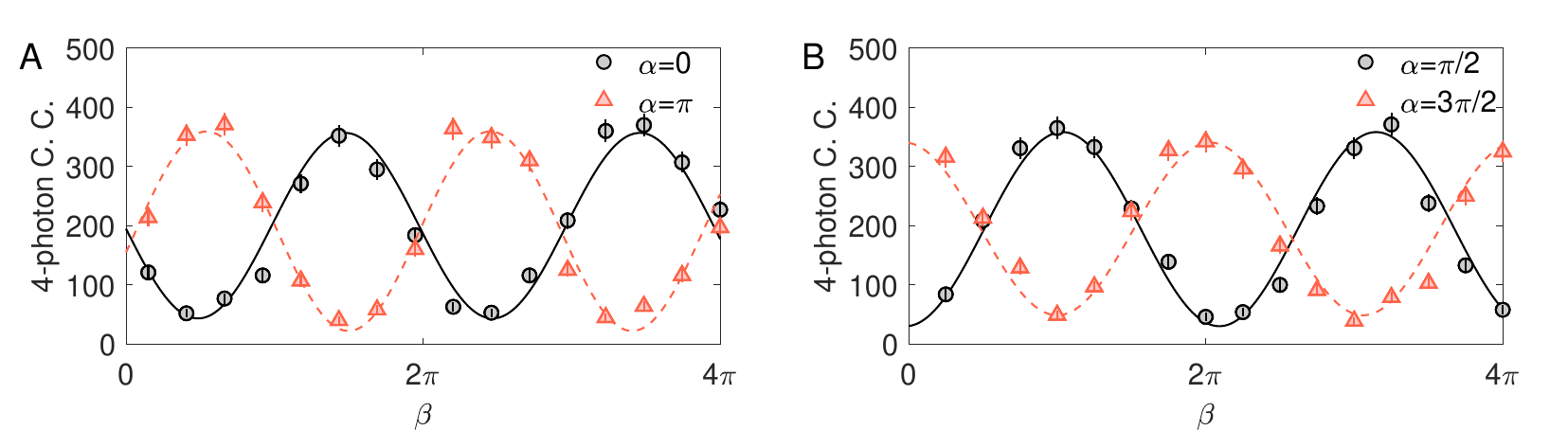}
\caption{Correlation results of four-photon-coincidence counts (C. C.) $N(\alpha, \beta)$. (A) The coincidence of Alice and Bob in fixed settings $\alpha = 0$ (black circles, fitted with black solid curve) and $\alpha= \pi$ (red triangles, fitted with red dashed curve). The visibilities of the two curves are $0.754\pm 0.021$ ($\alpha= 0$) and $0.805\pm 0.019$ ($\alpha= \pi$), respectively. (B) The coincidence of Alice and Bob in fixed settings $\alpha= \pi/2$ (black circles, fitted with black solid curve) and $\alpha= 3\pi/2$ (red triangles, fitted with red dashed curve). The visibilities of the two curves are $0.779\pm 0.020$ ($\alpha= \pi/2$) and $0.795±0.020$ ($\alpha = 3\pi/2$), respectively. The horizontal axis represents the settings of Bob ($\beta$), which varies from 0 to 4$\pi$. The error bars are derived from Poissonian distributions. Each data point is measured for 60 s.}
\end{figure*}
where $\mathcal{N}(\alpha,\beta)$ is the four-fold-coincidence count rate (Eq. (2), $N(+1,+1|\alpha,\beta))$ of $a_1$, $a_2$, $b_1$ and $b_2$ for the setting $(\alpha, \beta)$ of Alice and Bob. The probabilities of the other possible outcomes are:
\begin{widetext}
\begin{align}
  \nonumber p(-1,+1|\alpha ,\beta )&=\frac{\mathcal{N} (\alpha+\pi ,\beta )}{\mathcal{N} (\alpha ,\beta )+\mathcal{N} (\alpha +\pi ,\beta )+\mathcal{N} (\alpha ,\beta +\pi )+\mathcal{N} (\alpha +\pi ,\beta +\pi )}\\
  &=\frac{1}{4}-\frac{1}{4}\cos\mathrm{(}\alpha +\beta ),\\
  \nonumber p(+1,-1|\alpha ,\beta )&=\frac{\mathcal{N} (\alpha ,\beta+\pi )}{\mathcal{N} (\alpha ,\beta )+\mathcal{N} (\alpha +\pi ,\beta )+\mathcal{N} (\alpha ,\beta +\pi )+\mathcal{N} (\alpha +\pi ,\beta +\pi )}\\
  &=\frac{1}{4}-\frac{1}{4}\cos\mathrm{(}\alpha +\beta ),\\
  \nonumber p(-1,-1|\alpha ,\beta )&=\frac{\mathcal{N} (\alpha+\pi ,\beta+\pi )}{\mathcal{N} (\alpha ,\beta )+\mathcal{N} (\alpha +\pi ,\beta )+\mathcal{N} (\alpha ,\beta +\pi )+\mathcal{N} (\alpha +\pi ,\beta +\pi )}\\
  &=\frac{1}{4}+\frac{1}{4}\cos\mathrm{(}\alpha +\beta ),
\end{align}
\end{widetext}
This method can be viewed as an analogy to Bell experiments using polarizers (which exhibit $\pi$-period polarization rotation). In Bell experiments based on polarizers\cite{PhysRevLett.28.938,PhysRevLett.47.460}, when the polarizer is set to 0°, only the +1 (H) outcome is detected while the -1 (V) outcome is filtered out; conversely, when set to 90°, only the +1 (V) outcome is observed while the -1 (H) outcome is filtered out. For detailed comparison of constructing joint probabilities between entangled two photons (as in standard Bell experiment \cite{PhysRevLett.28.938,PhysRevLett.47.460}) and unentangled four photons by post-selection (as shown in this work), see Appendix. Therefore, once Alice and Bob choose their specific settings of phase $\alpha$ and $\beta$, the normalized expectation value of their result is:

\begin{align}
  \nonumber E(\alpha ,\beta )&=p(+1,+1|\alpha ,\beta )-p(-1,+1|\alpha ,\beta )\\
  \nonumber &-p(+1,-1|\alpha ,\beta )+p(-1,-1|\alpha ,\beta )\\
  &=\cos\mathrm{(}\alpha +\beta ).
\end{align}
With Eq. (7), we construct the Clauser–Horne–Shimony–Holt (CHSH) form of the Bell inequality \cite{PhysRevLett.23.880} for the four-photon FI:
\begin{equation}
  S=|-E(\alpha _1,\beta _1)+E(\alpha _1,\beta _2)+E(\alpha _2,\beta _1)+E(\alpha _2,\beta _2)|.
\end{equation}
Therefore, we have established a connection between our experimental configuration and the CHSH inequality through the joint probability distribution. It should be noted that in our scheme, the single observer measures the two-fold coincidence counts, rather than single counts as in the standard CHSH-Bell experiment. When we set $\alpha_1= 0, \alpha_2= \pi/2, \beta_1= \pi/4, \beta_2= 3\pi/4$, the Bell parameter S reaches the maximum value of $2\sqrt{2}$. Note that the classical bound of the S-parameter is 2.

\section{Results}
Experimentally, we measure the quantum correlation between Alice and Bob by fixing $\alpha$ at $\alpha_1= 0, \alpha_1^\bot= \pi, \alpha_2= \pi/2, \alpha_2^\bot= 3\pi/2$ and sweeping the phase $\beta$. The results of the four-fold-coincidence counts are shown in Fig. 2. The four-fold-coincidence counts for two orthogonal bases with phases $\alpha= 0$ and $\alpha=\pi$, respectively, show complementary results (Fig. 2A). The other two orthogonal bases with $\alpha=\pi/2$ and $\alpha=3\pi/2$ also yield complementary results (Fig. 2B). For the systematic analysis of the complete data, see Fig. A3 and A4 in Appendix.

\begin{figure*}
\includegraphics[width=14 cm]{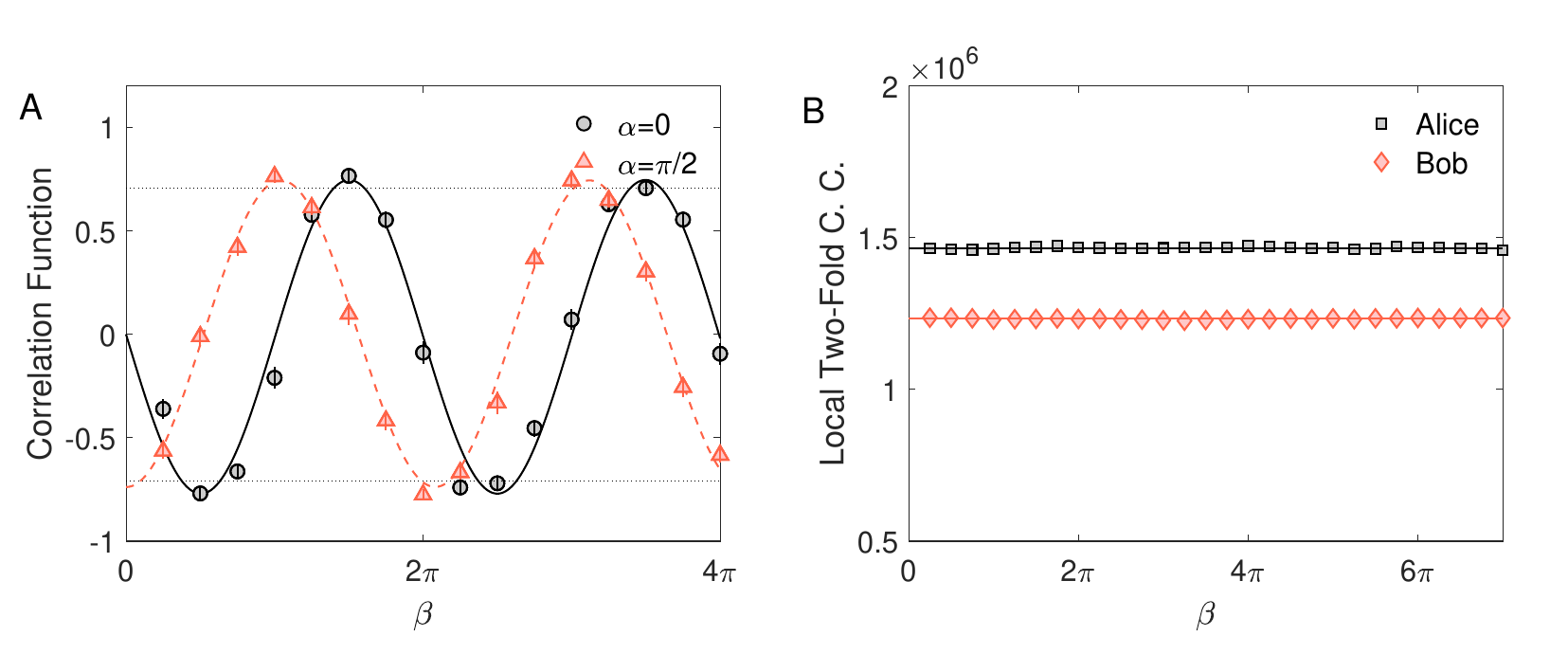}
\caption{Verification of quantum correlation. (A) Correlation functions $E(\alpha,\beta)$. The x-axis represents phase $\beta$. The maximum values of the correlation functions are $0.766±0.029$ ($\alpha= 0$) and $0.784±0.028$ ($\alpha = \pi/2$), respectively. (B) Local results of Alice and Bob. The non-local interference in four-photon FI shows no interference with only local detection.}
\end{figure*}

The correlation functions $E(\alpha,\beta)$ (Eq. 7) of Alice and Bob are shown in Fig. 3A, which are calculated from the data presented in Fig. 2. The maximum values of the correlation functions are 0.766±0.029 ($\alpha= 0$) and $0.784\pm0.028$ ($\alpha= \pi/2$), respectively, both of which are above the threshold ($V=1/\sqrt{2}$) as required for violating the CHSH inequality \cite{PhysRevLett.23.880}. We calculate the CHSH inequality from the coincidence counts (Fig. 2), which are listed in Table A1 in Appendix. We obtain a Bell parameter S of $2.275\pm 0.057$, which means that the Bell inequality is violated by more than four standard deviations. Note that in our previous work \cite{qian2023multiphoton}, the visibility and its statistical significance were not high enough for achieving confident violations of the Bell inequality. We have further systematically measured the four-fold FI over more periods. The detailed results are shown in Fig. A4. The visibilities are all above the threshold ($V=1/\sqrt{2}$). The observed violation of the CHSH inequality implies that the interference phenomenon we have witnessed cannot be accounted for by local realism. One may think the entanglement between vacuum and four-photon product states leads to the violation of CHSH inequality. In section A2 of Appendix, we show that the generated state from pair-creation processes contains a significantly smaller amount of entanglement than necessary for the observed Bell-inequality violation.

Fig. 3B shows the local two-fold coincidence counts of Alice ($\mathrm{C. C.} (a_1, a_2)$) and Bob ($\mathrm{C. C.} (b_1, b_2)$). The counts show no interference. This means the interference seen in Fig. 2 is not due to the local variation of Alice or Bob but arises from the correlation between the two parties. Importantly, Alice cannot derive the phase $\beta$ from her detection result. This is similar to measuring the local single counts of entangled states in a Bell test, which are independent of both Alice’s and Bob’s settings. 

\section{Discussion}
In the interferometer made by nonlinear crystals, we suppress or enhance the generation rate of the four-photon coincidence counts. Our results show that by using the measurement and the outcomes in the Bell-experiment formalism, we violate the CHSH inequality in the four-photon interferometer consisting of four non-linear crystals using path identity. In this work, indistinguishability of the creation processes — rather than entanglement — is responsible for our observation. Conceptually, our experiment differs from conventional Bell violation experiment in one crucial way: Rather than merely measuring an entangled quantum state, we actively manipulate the state during its creation. Remarkably, our experimental setup enables this manipulation to show quantum correlations via the violation of Bell inequality. 

One may have doubts about where the down-converted photons generated in sources I and II go when interference is destructive. In fact, the viewpoint that photons have been generated in sources I and II is incorrect. According to Bohr, “Physics concerns what we can say about nature.” \cite{petersen1963philosophy} If we would measure and can conclusively say that four photons have been generated by crystals I and II or III and IV, we obtain which-source information and therefore destroy the final interference. The right interpretation is that we merely create a possibility of generating four down-converted photons, rather than actually generating those four photons. Hypothetically, one could argue that we have created a multi-photon superposition between the two different four-photon sources (sources I\&II, and sources III\&IV after the post-selection of four-fold-coincidence counts), which leads to a multiphoton superimposed state $\ket{\psi}=1/\sqrt{2} [e^{i(\alpha+\beta)} \ket{1111}_{\mathrm{I\&II}}+\ket{1111}_{\mathrm{III\&IV}} ]$, where the subscripts represent the respective four-photon sources shown in Fig. 1A. However, this view is incorrect. As long as the photon paths are superimposed, the which-path information never exists \cite{PhysRevLett.67.318,RevModPhys.94.025007} and hence need not be erased, which is the fundamental difference between this work comparing to all existing quantum-eraser type works \cite{PhysRevA.25.2208,PhysRevLett.62.2209,PhysRevA.41.566,PhysRevLett.64.2495}. On the other hand, when all probability amplitudes corresponding to down-conversion emission processes leading to 4-photon coincidences interfere destructively, there is no consistent set of combinations of “physical” down-conversion emission processes that could lead to the detected 4-photon coincidence. Therefore, it is incorrect to argue that the state $\ket{\psi}$ is responsible for the Bell violation, since this process actually cannot happen when the interference is destructively. This is the actual “counter-intuitive” aspect of the work.

In this work, we post-select four-photon coincidence counts to violate the inequality, which could potentially be mitigated with highly efficient photon generation processes \cite{hubel2010direct}. We anticipate to address this limitation in future implementations. The setting events of $\alpha$ and $\beta$ could be in principle space-like separated in four-photon FI (see the scheme in Fig. 1). In the present work, the settings $\alpha$ and $\beta$ are close to each other, and hence the locality loophole remains open (alongside sampling loophole). Classical local realism theories have improved and found new loopholes over the past decades in the context of Bell-inequality violation with entangled photon pairs \cite{PhysRevLett.28.938,PhysRevLett.49.1804,PhysRevLett.81.5039,rowe2001experimental,scheidl2010violation,giustina2013bell,PhysRevLett.111.130406,PhysRevLett.112.110405}. We expect that tailored loopholes and local hidden varaible to the work reported here can be identified, but we also expect that they will be consistently excluded by hardware improvements of high-quality quantum photonic devices and experiments, as we witnessed in the ninety-year endeavor in the violations of local realism with entangled particles \cite{PhysRev.47.777,PhysRev.48.696,schrodinger1935gegenwartige,PhysicsPhysiqueFizika.1.195,hensen2015loophole,PhysRevLett.115.250401,PhysRevLett.115.250402,PhysRevLett.119.010402,PhysRevLett.121.080404}. Moreover, our work could very well lead to other interesting experiments, like in the development in Bell experiment. In analogy to Bell experiment with two particles, we expect quantum mechanics will finally prevail.

\section{Methods}
\subsection{Definition of probabilities}
We observe that the frustrated interference exhibits a 2$\pi$-period cosine dependence on the phase (see Fig. A3 in Appendix). In particular, the coincidence count under a $\pi$ phase shift, $N(+1,+1|\alpha+\pi,\beta)$, is complementary to the count in the original setting, $N(+1,+1|\alpha,\beta)$, and therefore can be identified as $N(-1,+1|\alpha,\beta)$, which is also complementary to $N(+1,+1|\alpha,\beta)$. Accordingly, we adopt the following relations between outcome counts and interferometer settings:
\begin{widetext}
\begin{align}
N(+1, +1|\alpha,\beta)= N(+1, -1|\alpha,\beta+\pi)= N(-1, +1|\alpha+\pi,\beta)=N(-1, -1|\alpha+\pi,\beta+\pi)\\
N(+1, -1|\alpha,\beta)= N(+1, +1|\alpha,\beta+\pi)= N(-1, -1|\alpha+\pi,\beta)=N(-1, +1|\alpha+\pi,\beta+\pi)\\
N(-1, +1|\alpha,\beta)= N(-1, -1|\alpha,\beta+\pi)= N(+1, +1|\alpha+\pi,\beta)=N(+1, -1|\alpha+\pi,\beta+\pi)\\
N(-1, -1|\alpha,\beta)= N(-1, +1|\alpha,\beta+\pi)= N(+1, -1|\alpha+\pi,\beta)=N(+1, +1|\alpha+\pi,\beta+\pi)
\end{align}
\end{widetext}
In general, the joint conditional probabilities should be defined in the same setting:
\begin{widetext}
\begin{equation}
  p\left( a,b|\alpha ,\beta \right) =\frac{N(a, b|\alpha ,\beta )}{N\left( +1,+1|\alpha ,\beta \right) +N\left( +1,-1|\alpha ,\beta \right) +N\left( -1,+1|\alpha ,\beta \right) +N\left( -1,-1|\alpha ,\beta \right)}
\end{equation}
\end{widetext}
where $a\pm 1$, $b\pm 1$, and we use $N(+1, +1|\alpha,\beta)$, $N(+1, -1|\alpha,\beta)$, $N(-1, +1|\alpha,\beta)$ and $N(-1, -1|\alpha,\beta)$ to construct the probability $p(a, b|\alpha,\beta)$. 
However, in our experiment, there is no -1 output. Therefore, based on Eqs (9)-(12), we use the following instead:
\begin{widetext}
\begin{equation}
  p\left( a,b|\alpha ,\beta \right) =\frac{N(+1, +1|\alpha +\frac{1-a}{2}\pi ,\beta +\frac{1-b}{2}\pi )}{N\left( +1,+1|\alpha ,\beta \right) +N\left( +1,+1|\alpha ,\beta +\pi \right) +N\left( +1,+1|\alpha +\pi ,\beta \right) +N\left( +1,+1|\alpha +\pi ,\beta +\pi \right)}
\end{equation}
\end{widetext}
Here, we use the terms which yield a (+1, +1) outcome and are accessible in our experiment. 

This approach is analogous to Bell experiments using polarizers \cite{PhysRevLett.28.938,PhysRevLett.47.460}, which exhibit a polarization rotation periodicity of $\pi$. Specifically, when the polarizer is set to 0°, only the +1 (H) outcome is detected, while the orthogonal -1 (V) outcome is blocked. These experiments typically rely on the assumption based on Malus’s law, the cosine dependence of the intensity of a polarized beam after an ideal polarizer \cite{groblacher2007experimental}. This assumption permits one to use the +1 outcome measured with the polarizer oriented at 90° to infer the otherwise inaccessible -1 outcome at 0°. In other words, by performing selective projective measurements along these mutually orthogonal directions, the combined outcomes form a complete projective measurement. We adopt the assumptions Eqs (9)-(12) when constructing the conditional probabilities, correlation functions and the inequality.

\subsection{The output state of the four-photon frustrated interference}
The unnormalized final state of the photons in modes $a_1$, $a_2$, $b_1$, $b_2$, truncated to second-order in the SPDC amplitude, is:
\begin{widetext} 
 \begin{align}
   \nonumber |\psi_f\rangle _{a_1a_2b_1b_2}&=|0000\rangle +g[e^{i\alpha}|1010\rangle +e^{i\beta}|0101\rangle +|1100\rangle +|0011\rangle]\\
   \nonumber &+g^2[e^{2i\alpha}|2020\rangle +e^{2i\beta}|0202\rangle +|2200\rangle +|0022\rangle ]\\
   \nonumber &+g^2[e^{i(\alpha +\beta )}|1111\rangle +|1111\rangle ]\\
   &+g^2\sqrt{2}[e^{i\alpha}|2110\rangle +e^{i\alpha}|1021\rangle +e^{i\beta}|1201\rangle +e^{i\beta}|0112\rangle],
 \end{align}
\end{widetext}
where each term denotes a Fock state, indicating the photon numbers in the respective modes $a_1$, $a_2$, $b_1$, $b_2$. The parameter g characterize the down-conversion gain (efficiency) of the SPDC process. We emphasize that Eq. (15) is an approximate expression. Notably, when evaluating the probabilities of local detection events observed by Alice, the calculated probability distribution $p_{a_1,a_2} (+1,+1)$ appears to depend on the phase $\beta$, which if naively interpreted might suggest superluminal signaling. This apparent paradox, however, arises from truncating higher-order terms in our theoretical framework. We confirm this through the following analytical derivations:
For each nonlinear crystal, we apply the two-mode squeezing operator:
\begin{equation}
  \hat{S}(g)=e^{-g(ab-a^{\dagger}b^{\dagger})}
\end{equation}
acting on input modes $a_{in}$ and $b_{in}$ \cite{RevModPhys.84.621}:
\begin{align}
  \left( \begin{array}{c}
	a_{\mathrm{out}}\\
	b_{\mathrm{out}}^{\dagger}\\
\end{array} \right) =\left( \begin{matrix}
	\cosh g&		\sinh g\\
	\sinh g&		\cosh g\\
\end{matrix} \right) \left( \begin{array}{c}
	a_{\mathrm{in}}\\
	b_{\mathrm{in}}^{\dagger}\\
\end{array} \right) 
\end{align}
From this, we obtain the four output modes of our experiment:
\begin{widetext}
\begin{align}
  \nonumber a_{1\mathrm{out}}&=\cosh ^2ga_{1\mathrm{in}}+\sinh ^2gb_{2\mathrm{in}}+\sinh g\cosh g(b_{1\mathrm{in}}^{\dagger}+a_{2\mathrm{in}}^{\dagger})
\\
\nonumber a_{2\mathrm{out}}&=\cosh ^2ga_{2\mathrm{in}}+\sinh ^2gb_{1\mathrm{in}}+\sinh g\cosh g(a_{1\mathrm{in}}^{\dagger}+b_{2\mathrm{in}}^{\dagger})
\\
\nonumber b_{1\mathrm{out}}&=\cosh ^2gb_{1\mathrm{in}}+e^{i\beta}\sinh ^2ga_{2\mathrm{in}}+\sinh g\cosh g(a_{1\mathrm{in}}^{\dagger}+e^{i\beta}b_{2\mathrm{in}}^{\dagger})
\\
b_{2\mathrm{out}}&=e^{-i\beta}\cosh ^2gb_{2\mathrm{in}}+\sinh ^2ga_{1\mathrm{in}}+\sinh g\cosh g(b_{1\mathrm{in}}^{\dagger}+e^{-i\beta}a_{2\mathrm{in}}^{\dagger})
\end{align}
\end{widetext}
where have we set $\alpha=0$. When Alice performs coincidence detections on paths $a_1$ and $a_2$, the second-order correlation is independent of $\beta$:
\begin{widetext}
\begin{equation}
  \langle a_{1\mathrm{out}}^{\dagger}a_{1\mathrm{out}}a_{2\mathrm{out}}^{\dagger}a_{2\mathrm{out}}\rangle =6\sinh ^4g\cosh ^4g+\sinh ^2g\cosh ^6g+\sinh ^6g\cosh ^2g
\end{equation}
\end{widetext}
Numerical simulations further validate this result (see section A1 in Appendix).

\section{Acknowledegement}
We thank \v{C}. Brukner for helpful discussions. This research was supported by the National Key Research and Development Program of China (Grants Nos. 2022YFE0137000), the Natural Science Foundation of Jiangsu Province (Grants Nos. BK20240006, BK20233001)), the Innovation Program for Quantum Science and Technology (Grants Nos. 2021ZD0300700 and 2021ZD0301500), Nanjing University-China Mobile Communications Group Co.,Ltd. Joint Institute, Jiangsu Funding Program for Excellent Postdoctoral Talent (No. 20220ZB60), and National Natural Science Foundation of China (Grant no. 12304397).

\clearpage

\section{Appendix}
\appendix
\renewcommand{\theequation}{A\arabic{equation}} 
\renewcommand{\thefigure}{A\arabic{figure}} 
\renewcommand{\thetable}{A\arabic{table}} 
\setcounter{equation}{0} 
\setcounter{figure}{0} 
\setcounter{table}{0} 

\section{A1. Numerical validation of the output state in the four-photon frustrated interference}
The numerical results, based on the experimentally obtained squeezing parameter g=0.096, are presented in Fig. A1(A). When the parameter g is sufficiently small, we perform a higher-order expansion of the squeezing operator S(g) with respect to g. As we gradually include higher-order terms of g, the local detection on Alice's side (coincidence on $a_1$ and $a_2$ when changing the phase $\beta$) exhibits a decreasing trend in visibility, which approaches zero. Therefore, under strict conditions, the visibility should asymptotically approaches to zero, and superluminal effects do not exist. Furthermore, the visibility of four-fold coincidence remains high Fig. A1(B). 

\section{A2. The analysis of entanglement between vacuum and four-photon product states}
Here we show that our result on the observed Bell-inequality violation cannot be explained by the entanglement between vacuum and four-photon product states. Consider the following bipartite state for Alice and Bob:
\begin{equation}
  |\psi _f\rangle _{AB}=C_g[|00\rangle _A|00\rangle _B+g^2(1+e^{i(\alpha +\beta )})|11\rangle _A|11\rangle _B]
\end{equation}
Where $C_g$ is the normalized coefficient. We define $\ket{00}_{AB}$ and $\ket{11}_{AB}$ of as the two-particle quantum states possessed by Alice/Bob. The parameter g is the efficiency of the spontaneous parametric down conversion process and we experimentally obtain its value of $g=0.096\pm0.008$. Without the loss of generality, we set $\alpha$ and $\beta$ to 0. We compute the value of Bell inequality, and find that $S_{\mathrm{vac}}=1.467\pm0.009$ \cite{PhysRevLett.23.880}. This value is below the bound of Bell inequality ($S_{\mathrm{classical}}=2$), and significantly below our measured value of $S_{\mathrm{exp}}=2.275\pm0.057$. Therefore, the entanglement between vacuum and four-photon product states cannot explain our result.

\section{A3. The detailed experimental setup and the systematic analysis of the complete data}
The detailed experimental setup of the four-photon frustrated interferometer is shown in Fig. A2. Fig. A3 shows the complete data of four-fold coincidence count as in Fig. 2 in the main text. The exact counts at the specific settings for violating Bell’s inequality is shown in Table. A1. To fully characterize the four-photon FI, we have further systematically measured the four-fold coincidence counts over more periods of different phases $\alpha$ and $\beta$. The detailed results are shown in Fig. A4.

\section{A4. Comparisons between violation of Bell’s inequality with entangled two photons and unentangled four photons}
We firstly consider the standard experiment setting for the violation of Bell’s inequality with entangled two photons \cite{RevModPhys.86.419}. A source emits two polarization-entangled photons to distant observers, Alice and Bob. Upon receiving their respective photons, each observer performs a measurement on it with their own apparatus. The measurement chosen by Alice/Bob is labeled $\alpha$/$\beta$ and its outcome a/b. Then, Alice and Bob record and analyze their joint probability distribution of acquiring outcomes a and b when Alice and Bob choose measurements $\alpha$ and $\beta$, $P(ab|\alpha\beta)$.  If one uses polarizers to measure the polarization states of both photons, as shown in the work\cite{PhysRevLett.28.938,PhysRevLett.47.460}, one measurement setting input gives one outcome +1 or -1. By rotating their polarizers and uses four combinations of measurement settings, Alice and Bob obtain the required joint probabilities to violate Bell’s inequality. 

In the four-photon FI, we send unentangled four photons to Alice and Bob. They each have two photons. The measurement choices $\alpha$ and $\beta$ are the phase settings of Alice and Bob. The output +1 represents detecting the two photons simultaneously ($a_1$, $a_2$ for Alice and $b_1$, $b_2$ for Bob, see Fig. 1 in the main text). Having one input at Alice or Bob’s side would produce one output result +1. We set an orthogonal base $\alpha+\pi$ to get the result -1 when we detect two photons simultaneously in the base $\alpha+\pi$. That means we adopt the fair-sampling assumption to ensure the validity of our results. Also, we assume $P(-1,+1|\alpha,\beta)=P(+1,+1|\alpha+\pi,\beta)$, this is from the definition of our measurement. In this way, one can see the measurement-setting and single-outcome correspondence between our work and standard Bell experiment with polarizers. However, we emphasize that here we use unentangled four photons with post-selection.

\begin{figure*}
\includegraphics[width=14 cm]{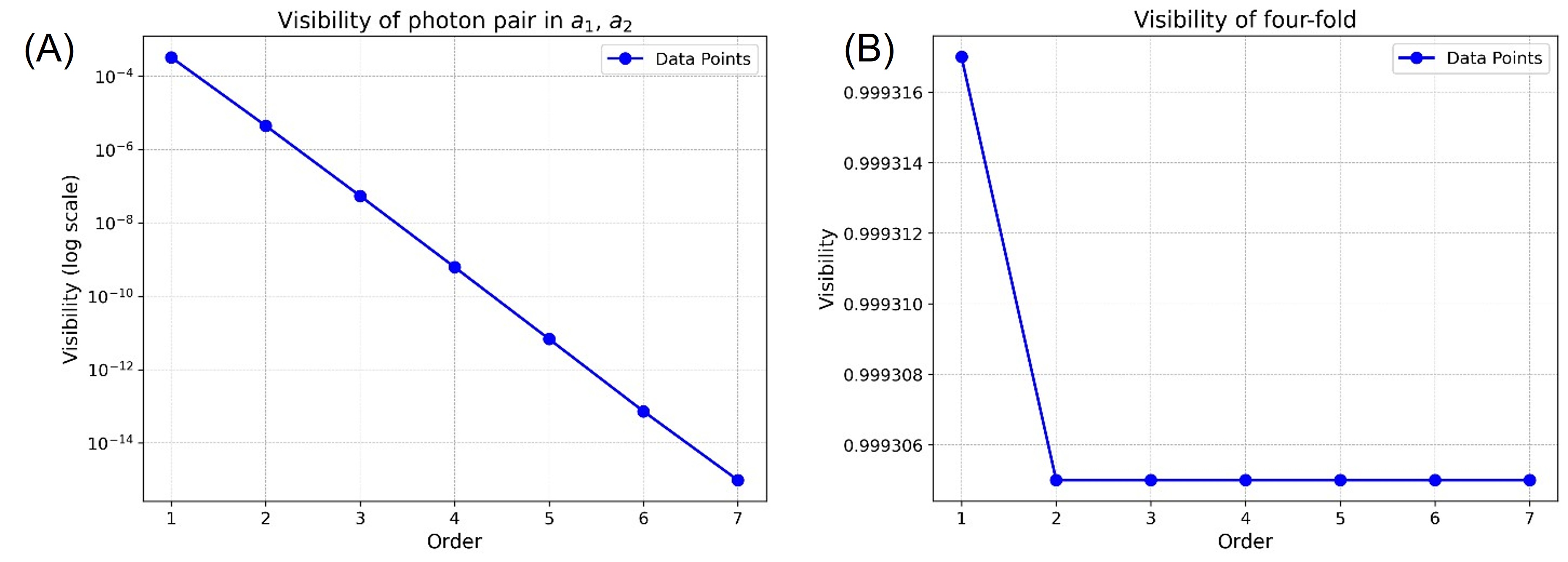}
\caption{Numerical Validation of Causality Preservation. (A) Dependence of two-fold coincidence visibility on squeezing operator order: The interference visibility of coincidence on $a_1$ and $a_2$ ultimately vanishing when higher-order terms of the squeezing operator are systematically incorporated. (B) In contrast, the four-photon coincidence visibility persists despite the inclusion of higher-order squeezing operator components.}
\end{figure*}

\begin{figure*}
\includegraphics[width= 14 cm]{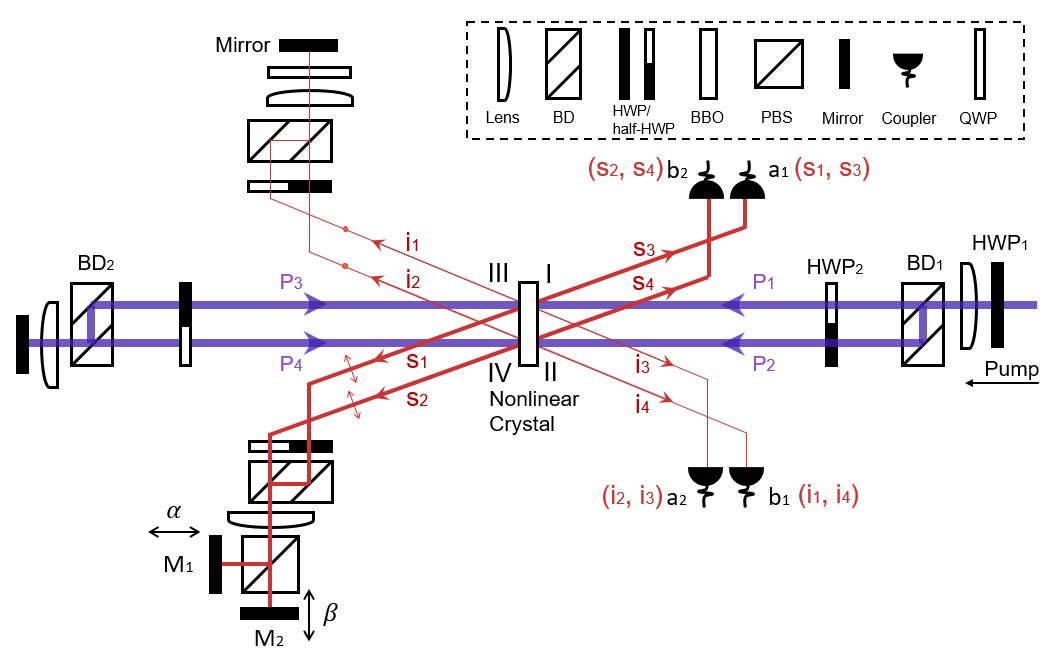}
\caption{Detailed experimental setup. From the right side, the pump beam incidents upon a beam displacer ($\mathrm{BD}_1$), splitting into two components. These components undergo spontaneous parametric down-conversion (SPDC) in a 'back-reflect' configuration, generating four photons. The idler photons from sources I and II (denoted as $i_1$ and $i_2$) undergo a polarization-based path exchange and are reflected by a mirror. The phases of the signal photons, $\alpha$ and $\beta$, are independently controlled by mirrors $\mathrm{M}_1$ and $\mathrm{M}_2$, respectively. To ensure path identity, $i_1$, $i_2$, $s_1$, and $s_2$ are aligned with $i_3$, $i_4$, $s_3$, and $s_4$, respectively. Finally, all four photons are collected by couplers $a_1$, $a_2$, $b_1$, $b_2$, and detected using single-photon detectors.}
\end{figure*}

\begin{figure*}
\includegraphics[width= 14 cm]{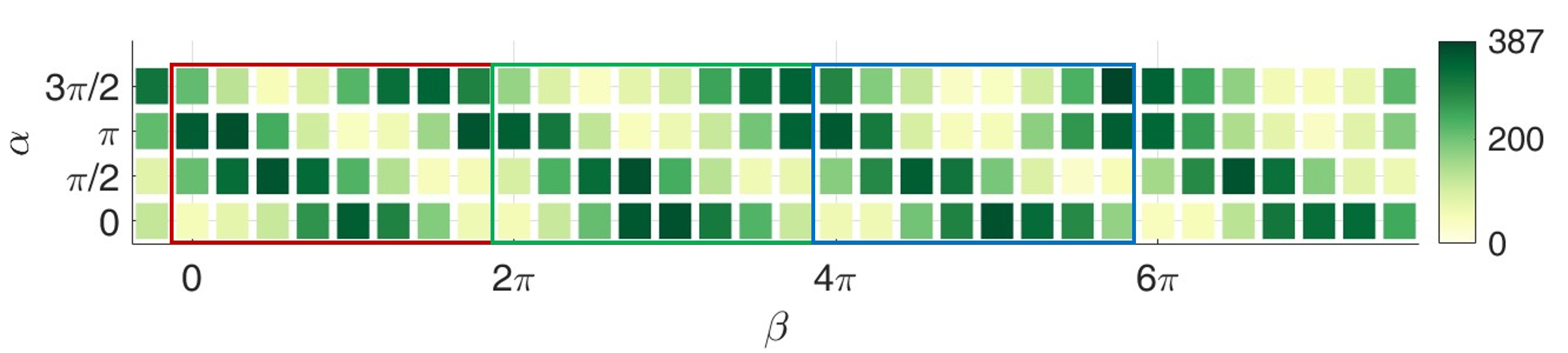}
\caption{Four-photon-coincidence count of different phase settings $\alpha$ and $\beta$. The raw data of four-photon-coincidence count is shown. For each row, we set $\alpha$ for different values (0, $\pi/2$, $\pi$, $3\pi/2$) and sweep the phase $\beta$ for four periods of $2\pi$. The integration time of each data point is 60 seconds. The red box indicates the data used to calculate the CHSH parameter S in the main text. The parameter S calculated from all the data above (red + green + blue box) is $S=2.106\pm0.046$.}
\end{figure*}

\begin{figure*}
\includegraphics[width= 14 cm]{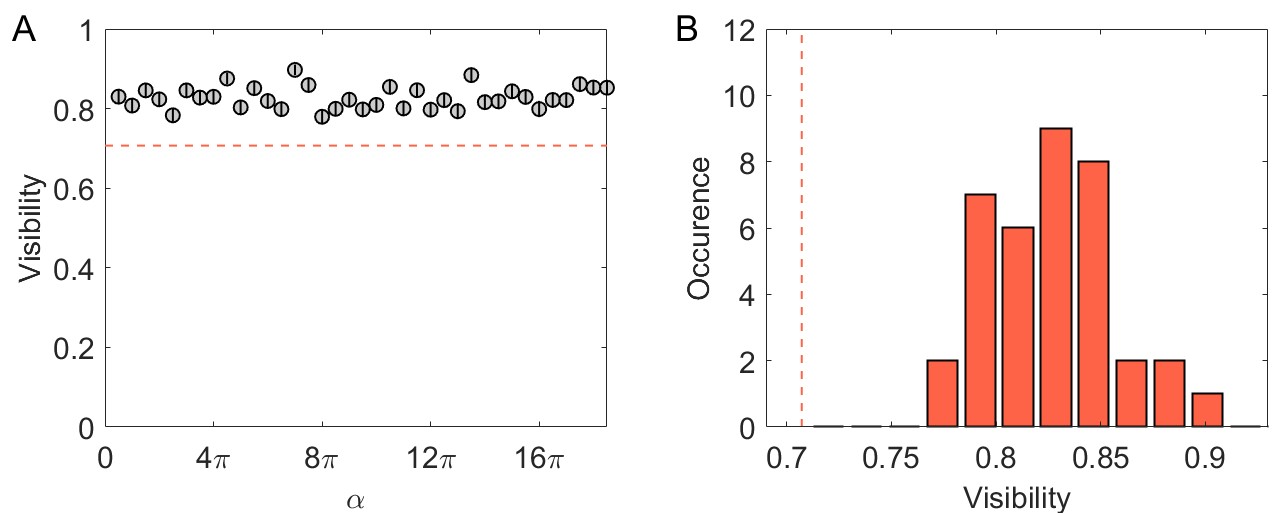}
\caption{The visibility of the four-photon interference using quantum indistinguishability by path identity. (A) We set a series of phase points for $\alpha$, ranging from 0 to $18\pi$ with increments of $\pi/2$ and again sweep the phase $\beta$ over four periods of $2\pi$. For each distinct phase $\alpha$, we determine the visibility and associated error bar from the maximum and minimum counts observed at the specific phase $\alpha$ and error propagation. The overall visibilities are above the threshold ($V=1/\sqrt{2}$, red dot line) set by the CHSH Bell inequality. (B) The statistics of visibilities for each point in Fig. A4(A) shows an interference with visibility always above the threshold $1/\sqrt{2}$. The average visibility through all phases $\alpha$ is $V=0.828\pm0.018$. By assuming the general white noise and hence the resulting Werner state form, we obtain the Bell violation of $S=2.342\pm0.051$ based on S-V relation. Our results show the conflict between quantum mechanics and local realism using quantum indistinguishability.}
\end{figure*}

\begin{table}[htbp]
  \centering
  \caption{Four-photon-coincidence counts (60 s measurement time)}
  \label{tab:alpha-beta-counts}
  \begin{tabular}{|c|c|c|c|c|c|c|}
    \hline
    \multicolumn{3}{|c|}{} & \multicolumn{4}{c|}{$\beta$} \\ \cline{4-7}
    \multicolumn{3}{|c|}{} & $\beta_1$ & $\beta_1 + \pi$ & $\beta_2$ & $\beta_2 + \pi$ \\ \cline{4-7}
    \multicolumn{3}{|c|}{} & $\pi/4$ & $5\pi/4$ & $3\pi/4$ & $7\pi/4$ \\ \hline
    \multirow{4}{*}{$\alpha$}
      & $\alpha_1$       & 0 &  77 & 295 & 271 &  63 \\ \cline{2-7}
      & $\alpha_1 + \pi$ & $\pi$ & 371 &  58 & 107 & 364 \\ \cline{2-7}
      & $\alpha_2$       & $\pi/2$ & 331 & 139 & 333 &  54 \\ \cline{2-7}
      & $\alpha_2 + \pi$ & $3\pi/2$ & 129 & 327 &  97 & 296 \\ \hline
  \end{tabular}
\end{table}

\clearpage
%

\end{document}